\documentstyle[aps,epsfig]{revtex}
\begin{document}
\draft
\title
{Second harmonic generation and birefringence of some ternary pnictide
semiconductors}
\author{Sergey N.~Rashkeev,\cite{a} Sukit Limpijumnong 
and Walter R. L. Lambrecht}
\address{Department of Physics, Case Western Reserve University, Cleveland,
OH 44106-7079}
\date{\today}
\maketitle

\begin{abstract}
A first-principles study of the birefringence and the frequency dependent 
second harmonic generation (SHG) coefficients
of the ternary pnictide semiconductors with formula ABC$_2$ 
(A = Zn, Cd; B = Si, Ge; C = As, P) 
with the chalcopyrite structures 
was carried out.  It uses a recently developed computational approach
based on the self-consistent linear muffin-tin orbital (LMTO) 
band-structure method, which is applied using the local density
approximation to density functional theory with a simple {\em a-posteriori}
gap correction. The susceptibilies are obtained 
in the independent particle approximation, i.e., without local field 
and excitonic effects.
The zero frequency limits of $\chi^{(2)}_{123}$
were found to be in reasonable agreement
with available experimental data for all the considered materials.
We found that substitution of P by As, Si by Ge, and Zn by Cd is
favorable to get a higher value of $\chi^{(2)}(0)$.
However, the anomalously high value of the zero frequency SHG in
CdGeAs$_2$ (the material with the most favorable combination
of A, B, and C) is rather exceptional than typical for this group of compounds.
An analysis of the different contributions in the frequency dependent
SHG spectra shows that
this value appears as a result of a very small interband
term in the zero frequency limit which cannot compensate
the large intraband contribution as it happens in most of the other 
materials of this class. The smallness of the interband term in
CdGeAs$_2$ is a result of a very delicate balance
between different interband transitions, any of which
can give positive or negative contribution in SHG.
While we find the empirical observation that 
a smaller value of the gap is strongly correlated
with a larger value of the SHG to be generally true when 
comparing various element substitutions within this family 
in a qualitative sense, simple inverse power scaling laws 
between gaps and $\chi^{(2)}$ values are not supported by our 
results.  The case of CdGeAs$_2$ 
clearly shows that this is an oversimplification and that for reliable 
predictions of trends of SHG coefficients  
one has to study the interplay between different terms 
which contribute in SHG. We have also studied the relation between 
$\chi^{(2)}$ and the chalcopyrite crystal structure by considering some 
of these materials in an alternative layered zincblende type latttice.
We find that the (001) oriented $1+1$ superlattice structure 
has significantly lower gaps than
the chalcopyrite and correspondingly higher $\chi^{(2)}$. 
However, this smaller gap structure is characterized by
a large alternatingly compressive and tensile 
lateral strain in the layers, which makes it unfavorable.
We also find that the distortions from the ideal chalcopyrite tend 
to increase the gap and decrease both the inter- and  intraband
contributions to $\chi^{(2)}$, but the net value of $\chi^{(2)}$
is only slightly changed in most cases.  
These effects  are larger for the Cd compounds than 
for the Zn-compounds. 
As far as the birefringence is concerned, we find 
our calculations for ZnGeP$_2$ and CdGeAs$_2$ to be 
in fair agreement with experiment 
(discrepancies being rather constant and of order 10 \%) 
in the frequency range 
corresponding to the middle of the gap but 
to deviate from the data when the absorption edges
of the band gap at high energy and the phonon absorption 
bands at low energy are approached.
It is presently not clear whether this reflects a deficiency 
in the theory or imperfections in the samples.
\end{abstract}

\pacs{PACS numbers: 42.65.Ky, 78.20.Bh }

\section{Introduction} \label{sec:int}

Ternary chalcopyrites  are promising  for optical frequency 
conversion applications 
in solid state laser systems, such as optic parametric
oscillators (OPO) and frequency doubling devices \cite{boss}.
Zinc germanium diphosphide (ZnGeP$_2$) is an excellent nonlinear
optical material which exhibits good optical transparency
over the 0.7--12 $\mu$m wavelength region.
This uniaxial crystal has positive birefringence, and its
conventional conversion efficiency, or
figure of merit (FOM=$(\chi^{(2)})^2/4n^3$ where $n$ 
is the index of refraction)
exceeds that of LiNbO$_3$, a presently commonly used
material, by an order of magnitide. It is therefore,
a very good candidate to producing tunable laser output
in the near infrared \cite{Zn-Ohm1,Zn-Ohm2,ZGP1,ZGP2,ZGP3}.
CdGeAs$_2$ has even larger $\chi^{(2)}$ and FOM. 
To the best of our knowledge
this material has the highest nonlinear optical coefficient 
in the class of phase-matchable compounds \cite{Cd-Ohm,CGA1,CGA2,Dmitriev}.
Recent crystal growth technologies have made great progress towards growth of  
large crystals of both ZnGeP$_2$ and CdGeAs$_2$ with improved optical
quality. These two materials, however, are part of a larger family of 
II-IV-V$_2$ compounds and although the crystal growth 
perfection of these has not been
pursued to the extent of the above two, it is of interest 
to ask whether other materials 
in this class, or alloys between them might offer increased flexibility 
in terms of combining high $\chi^{(2)}$
with other ranges of transparency and/or 
non-critical phase-matching.
More generally, it is of interest 
to understand the origin of the high $\chi^{(2)}$ and
the degree of birefringence in these materials
as well as to study the trends 
within this family of compounds.

In the present paper we systematically study the electronic 
structure and optical
properties of a class of ternary pnictides with formula  
ABC$_2$ (A = Zn, Cd; B = Si, Ge; C = As, P) 
which are crystallizing in the chalcopyrite structure.
The measured SHG coefficients have been reported for some of these materials,
and their values are generally rather high (at least not much lower
than the SHG in GaAs) \cite{Dmitriev,Landolt}. Our main purpose is to 
understand the trends in this family of materials and to check whether 
simple schemes for extrapolation (e.g., based on inverse 
power relations to the band gap)
to other compounds in this family are valid. 
The choice of A, B, and C atoms allows us 
to consider the trends in the group-II, group-IV, and group-V atoms separately 
in this class of materials which has the general formula II-IV-V$_2$.
We also note that some of these materials have direct gaps while others have 
indirect or pseudodirect
gaps. Thus, we can investigate whether there is any relation 
between the type of the gap
and the $\chi^{(2)}$ values. In fact, we find little evidence for 
such a correlation. This result could be expected from 
the fact that $\chi^{(2)}$ 
depends on a Brillouin zone average of interband vertical optical transitions 
and is, therefore, little affected by the bands at a few particular 
{\bf k}- points 
determining the band edges. 

Typically, one indeed finds smaller band gap materials
to have higher $\chi^{(2)}$ \cite{chalcopyrites}.
It is expected from the basic expressions for the NLO response functions 
\cite{Sipe,Aversa} and from empirical observations that NLO susceptibilities
scale roughly inversely with some power of the interband energy differences,
and, therefore, with the minimum gaps if these dominate the response.
However, we show
that the situation is not that straightforward, and the real value
of the zero frequency limit of SHG is a result of a very delicate
balance between different contributions to nonlinear optical response,
e.g., the intraband and interband contributions. It turns out that 
CdGeAs$_2$ is a rather exceptional material in this context.

A second purpose is to investigate whether or not the 
chalcopyrite structure plays a 
special role in leading to high $\chi^{(2)}$ values. 
We can test the effect of crystal 
structure at least qualitatively by considering an alternative  
crystal structure corresponding to the same overall 
chemical formula or stoichiometry.
To this end we consider a (001) oriented $1+1$ 
superlattice structure and compare its 
band structure, total energy, and $\chi^{(2)}$ 
with that of chalcopyrite for ZnGeP$_2$.
We note that this particular structure 
still maintains the local tetrahedral environment
of the chalcopyrite corresponding to each anion being 
surrounded by two cations of each type.
Our results show that the alternative structure 
considered generally has a much lower band gap
and hence higher $\chi^{(2)}$. 
However, we find that this is related to the 
occurence of alternatingly compressive and tensile
biaxial strain in the layers which make 
the structure unstable. Thus, we do not think that this alternative 
structure gives a promise for further increases of $\chi^{(2)}$. 

The chalcopyrite structure is typically accompanied 
by a distortion from the ideal $c/a$ ratio 
one would obtain by simply substituting II and IV 
atoms in the cubic zincblende structure 
in the appropriate ordering. It is thus also 
of interest to investigate whether this has an important 
effect on $\chi^{(2)}$. We find that these effects are 
significantly larger  for the Cd than for the Zn compounds 
which are closer to the ideal structure. While the gaps 
are generally increased and hence, both intra- and interband contributions
are significantly decreased, the net value of $\chi^{(2)}$ is only 
slightly changed  as a result of the compensation noted above.

The computational approach used in the present 
investigation was described in
our previous paper\cite{Our1}
and successfully applied to SiC polytypes \cite{Our2}
and GaN/AlN superlattices \cite{Our3}. 

The rest of the paper is organized as follows. 
In Section \ref{sec:comp} we describe briefly 
the necessary details of our computational approach. 
In Section \ref{sec:res} we present our calculated results for the electronic 
band structures, static second harmonic generation 
coefficients and birefringence 
and address the various questions raised 
in this introduction. We analyze the results  
in terms of the decomposition of the frequency 
dependent Im$\chi^{(2)}(-2\omega,\omega,\omega)$
function on intra- and interband terms as defined 
in Refs. \onlinecite{Sipe} and \onlinecite{Aversa}. 
A summary of the main conclusions of this work is given in Sec. ~\ref{sec:con}.

\section{Computational Method} \label{sec:comp}

Our calculations are based on the 
linear muffin-tin orbitals method (LMTO)
\cite{OKA,lmto} within the atomic sphere approximation (ASA).
This method is quite efficient mainly because 
it employs a rather small basis set.
It allows to easily deal with systems
containing a large number of atoms per unit cell 
while maintaining a sufficiently large number of 
{\bf k}-points so as to ensure converged Brillouin zone (BZ) integrations.
Extensive checks performed in our 
previous paper \cite{Our1} demonstrated that our ASA-LMTO based approach
yields results for the second
order response functions of comparable accuracy as the full-potential 
linear augmented plane-wave method (FLAPW) method\cite{Hughes,Hughes1}.
The formulas used here to calculate $\chi^{(2)}$
were given in a previous publication.\cite{Our1} They are based
on the so-called ``length-gauge'' formalism of 
Sipe and Ghahramani,\cite{Sipe} and Aversa and Sipe\cite{Aversa}.
This formalism uses the independent particle
approximation and is presently restricted to undoped semiconductors.
Also, it does not include the local field effects.
This method has several apparent advantages, e.g.,
the manifest absence of unphysical 
singularities of nonlinear optical responses in the zero-frequency
limit and the explicit  satisfaction of the Kleinman relations.\cite{Kleinman}

The self-consistent LMTO electronic structure calculations
were carried
out within the framework of the local density approximation (LDA) in
the density functional theory.
The exchange-correlation potential has been taken in the form of
Hedin and Lundqvist.\cite{Hedin}
One of the central problems in calculation of optical response functions
in semiconductors is the problem of the gap corrections in the LDA
\cite{Our1,Hughes,Hughes1,Allan,Chen} which
appears from the fact that response functions have to deal with
the actual quasiparticle excitations rather than the
Kohn-Sham eigenstates.
The presently  most accurate approach, the GW method,
is unfortunately rather computationally demanding 
because it needs calculations of 
matrix elements of the nonlocal
self-energy operator in terms of the one particle Green's function and the 
screened Coulomb interaction which requires itself calculation of 
the inverse dielectric response function 
including local field effects.\cite{HedinGW}
In practice, a simplified approach, the so-called ``scissors operator''
is often used. It assumes the rigid shift in energy
$\Delta$ of the conduction band with subsequent renormalization of
the velocity (momentum) operator matrix elements
(see, e.g., Refs.\onlinecite{Our1,Hughes,Allan} for details).
In our previous paper \cite{Our1} we noted that this approach is not entirely 
satisfactory because it breaks the consistency between
the eigenvectors and eigenvalues of the one 
particle Hamiltonian. This manifests itself 
rather clearly when applying this approach 
to the calculation of effective masses. In a rigid shift,
the masses should not change, but 
applying the ``scissors'' renormalization to the momentum
matrix elements and using the well-known ${\bf k}\cdot{\bf p}$ expression for
the inverse masses we see that the effective mass decreases.

We proposed an alternative approach, which consists 
in adding semi-empirical corrections 
to the diagonal elements of the LMTO. In some sense, it is an attempt of 
describing the scissor's operator $\sum_{ck}\Delta_{ck}
|ck\rangle\langle ck|$, in which $|ck\rangle$ 
are the conduction band states and 
$\Delta_{ck}$ their shifts,  
explicitly at the level of the Hamiltonian. 
The approach is based on the observation 
that the states at the bottom of the conduction band in tetrahedrally bonded 
semiconductors have a characteristic predominance 
of certain LMTO basis orbitals 
in their wave functions: e.g., cation $s$- at $\Gamma$ 
and a mixture of cation $s$- and 
empty sphere $s$- orbitals at the $X$- point of the zincblende BZ. 
Therefore, by 
shifting these $s$- orbitals, one shifts the corresponding states. 
The states at 
$L$ typically behave intermediately between those at $\Gamma$ and $X$. 
Thus, an advantage is that one can include {\bf k}- dependent 
shifts if these are known.  
This alternative approach is closely related to the one suggested
by Christensen \cite{Christensen} in which it is further 
used that only $s$- states 
are sensitive to sharply localized $\delta$-function-like potentials.
In our previous paper \cite{Our1} we showed that better results
for the frequency dependent SHG (at least for GaAs and GaP) 
can be obtained in this way.
However, this procedure allows us only to shift the lowest conduction
bands while for some materials the shift of the whole set of conduction
bands is a better approximation (e.g., for SiC, 
see Ref.\onlinecite{Our2} for details). 
In other words, it is a reasonable approach 
only if one is interested in the highest 
valence band and the lowest conduction band 
exclusively. Its success for GaAs and GaP observed in our previous work 
is thus based on the fact that these bands tend to dominate the second order 
response functions in that material but this is not a general fact 
for all semiconductors. 

In the present study, in order to study trends, it is important 
to follow a systematic and consistent treatment for all the compounds, 
including those for which detailed band gap information at different 
{\bf k}- points is not available.
Therefore, the bulk of our results was obtained using  the simple ``scissors
operator'' approach. Unfortunately, we cannot use that approach directly for 
the case of CdGeAs$_2$ 
(where LDA does not give any gap at all). 
As a result, the bands 
in CdGeAs$_2$ have a completely changed topology near the ``overlap'' of the 
valence and conduction bands at $\Gamma$. 
To deal with this case we first applied a shift of the Hamiltonian 
$s$- states 
to get some ``bare minimum'' gap of at least 0.3 eV and 
subsequently applied the usual scissors operator to obtain the experimental 
gap values. 

We emphasize that the primary purpose of this paper is to study the chemical 
and structural effects on the electronic structure and $\chi^{(2)}$ 
functions. The above described problems with the gap corrections, although 
not handled to our entire satisfaction, are expected 
to only have a minor effect on the main conclusions of our work as long as 
a consistent treatment is followed.  

The detailed description of the procedure for calculating the linear
response and frequency dependent SHG has been described elsewhere
(see, e.g., Ref.\onlinecite{Our1}).
The real part of the frequency dependent 
response functions has been obtained from the imaginary part by using the
the Kramers-Kronig transformation.
For the zero-frequency limit of SHG we use the simplified and computationally 
more convenient formulas.  
It avoids the need for calculation of the response function in a wide
frequency region required for an accurate application 
of the Kramers-Kronig transformation
and is less singular as regards {\bf k}- point summations, 
so it can be converged with 
fewer terms.
The {\bf k}- space integration
may be  limited to the irreducible wedge of the Brillouin zone 
by performing  a 
prior symmetrization of the product of the
three  momentum matrix elements over all the transformation of the 
point group of the crystal.

For the frequency dependent SHG, we use the usual tetrahedron scheme 
of the integration with linear interpolation of the band energies and the
products of the matrix elements. 
For the zero-frequency limit on the other hand,
we employ a semi-analytical linear
interpolation scheme which is more efficient and produces a smaller 
error.\cite{Our1}

\section{Results} \label{sec:res}

\subsection{Crystal structure and electronic bands} \label{ss:bnd}

The chalcopyrite ABC$_2$ materials (with chemical composition
II-IV-V$_2$ and I-III-VI$_2$)
can be obtained from III-V compound and II-VI
compounds respectively by replacing every two group III (II) atoms per cell
by a II and IV (I and III) atom. For example, from GaP we derive ZnGeP$_2$
by this chemical substitution. 
Their crystal structure is a body-centered tetragonal lattice which
has eight atoms per unit cell. It can be thought of as a tetragonally
distorted A$_2$B$_2$ (or 2 + 2) face centered cubic (fcc) superlattice
in the \{201\} direction of the cations A and B with an interpenetrating
fcc lattice of common anions C displaced by (1/4,1/4,$u$). The structural
parameters are the lattice constant $a$ which corresponds to the cubic
lattice constant of the zincblende structure from which the chalcopyrite 
structure is derived, the ratio $\eta=c/a$, and the internal displacement
parameter $u$. In the ideal structure $\eta$ = 2 and $u$ = 1/4.
The nonideal value of $u$ is due to the distortion of the anion sublattice
involving a shift by each anion away from one neighboring cation
in the direction of the another cation (of different sort). 
It was shown that such a shift of anions is a consequence
of atomic sizes \cite{Zung1,Zung2,Zung3}. Usually the column-II atom
is larger than the column-IV one, and the anions are closer to B than
to A atoms. 

In principle, one can obtain the $a$, $\eta$ and $u$ parameters
by total energy minimization. 
It was shown in a previous paper  
by one of the authors that such a relaxation can significantly affect 
the band structure \cite{chalco}.
We did not do such a relaxation in the 
present work  but preferred to use the available experimental 
values of these structural parameters. Some comparisons to the ideal structure 
were also carried out.
Table \ref{tab:str} shows experimental values of the
lattice parameters $a$, $\eta$, and $u$ for several chalcopyrites
ABC$_2$ from the considered group. 
The most complete set of parameters
can be found in Refs.\onlinecite{Wyckoff}
and \onlinecite{Landolt2}. 
The early measurements of Ref.\onlinecite{Wyckoff}
did not pay too much attention on detailed measurements
of the parameter $u$ but 
the measured values of the parameters $a$ and $\eta$ 
are very close in both the references.
Below we show that
the distortions from the ideal chalcopyrite structure 
($\eta$=2, $u$=1/4) can be neglected only when they are small.
Otherwise, they can 
change significantly the value of the gaps and the SHG. 
It would be interesting to mention that
the substitution of phosphorus by arsenicum increases the lattice
constant $a$ by the factor of 3.5--4 \% while the value of
$\eta=c/a$ changes by less than 1 \%. The substitution of Ge by Si
does not change either $a$ or $\eta$ by more than 1.5 \%.

Table \ref{tab:shg} shows the experimental values of the minimum 
band gaps in these
materials from Ref.\onlinecite{Landolt2} compared to 
our LDA gaps and their difference in the first three columns. 
The nature of the band gaps varies from direct to pseudodirect or 
indirect. By pseudodirect is meant that the gap is nominally direct 
but corresponds to a weak, almost forbidden, optical transition.
In the chalcopyrites, this situation can occur because  the zone edge states
of the parent III-V zincblende compound at $X^{zb}$ are folded onto the
$\Gamma$-point in the chalcopyrite structure.
Thus, when the zincblende 
$X_{1c}$ state is lower than the direct gap at $\Gamma$ (which happens in GaP),
a pseudodirect gap is expected in a chalcopyrite corresponding to a small 
perturbation from  GaP. 
The correspondence between states in chalcopyrite notation 
and the states in zincblende is as follows: 
the conduction band minimum at $X$, $X_{1c}^{zb}\mapsto \Gamma_3$,
the next higher state at $X$, 
$X_{3c}^{zb}\mapsto \Gamma_2$ and $\Gamma_{1c}^{zb}\mapsto \Gamma_1$.
The latter is predominantly s-like while the other two have 
important p-contributions.
The optical matrix elements between the crystal-field splitted
valence band maxima $\Gamma_4$ or $\Gamma_5$, which are both p-like, 
and the $\Gamma_1$ conduction band state are large while the ones 
with $\Gamma_3$ and $\Gamma_2$ are small. By inspection of the band
dispersion, or the matrix elements, or the eigenvectors, one 
can easily identify the nature of the lowest conduction band
at $\Gamma$ as being $\Gamma_1$ or $\Gamma_3$ like.  Our calculated
gaps indicted as being pseudodirect in fact correspond to the $\Gamma_3$
minimum while the direct ones correspond to a $\Gamma_1$ minimum. 
The experimental assignments are mainly based on the 
optical absorption being weak or strong \cite{Sileika} along with 
some general theoretical considerations based on zone folding arguments
of the type explained above. Experimentally, it is very difficult 
to distinguish between a pseudodirect and a truly indirect gap.

Although all the LDA gaps are lower than the experiment by about 1 eV,
we note that the assignments of the direct or pseudodirect 
nature agree with experiment with two exceptions. 
We find ZnGeP$_2$ to be actually indirect 
with a conduction band minimum location at a point near to the folded 
$X$-point. This will be further discussed elsewhere \cite{Limpijumnongzgp}
and is still in agreement with the absorption measurements indicating a weak 
transition. 
Also, we find ZnSiAs$_2$ to be direct while experiment indicates a 
pseudodirect gap. We note however, that in this case, the 
$\Gamma_3$ pseudodirect gap as well as the lowest indirect gap 
in our calculation are only 0.1 eV above the 
direct $\Gamma_1$ gap. Errors of the order of 0.1 eV
in the ordering of the conduction band states are common, a well-known 
example being Ge, for which it was found that subtle effects such as 
core-polarization \cite{Shirley} need to be included to obtain the correct 
ordering.
 
We note that if we order these semiconductors according to increasing 
gap, the lower gap ones are direct, the higher gap ones are pseudodirect 
(or indirect) with the crossing taking place at about 1.7 eV, 
i.e., CdGeP$_2$ and ZnSiAs$_2$ being borderline cases. 
From the above discussion relating the bands to those of the parent 
III-V compounds, the pseudodirect    
situation is clearly expected in ZnSiP$_2$, because Zn is
in the same row of the periodic table as Ga and the Si s-state
occur higher than the Si-p states ($\Gamma_{2'}>\Gamma_{15}$).
Thus one expects the $\Gamma_1$ state to be mainly Zn- like, and a 
a small perturbation from GaP is expected. 
In ZnGeAs$_2$, on the 
other hand, we clearly expect a truly direct ($\Gamma_1$-like) 
gap because GaAs is direct. 
Next, we note that Cd being a heavier element has lower s-like states
than Zn and thus replacement of Zn by Cd will tend to make the gap direct
as opposed to pseudodirect. Ge also has a lower $\Gamma_1$ state than 
Si. Thus replacing
Si by Ge will also tend to make the gap direct. This is consistent with 
the borderline case CdGeP$_2$ being direct.
Also, in ZnGeP$_2$, we find the 
$\Gamma_1$ state slightly below the $\Gamma_3$ but nevertheless it
is indirect with a minimum gap at another {\bf k}- point. Again, in this 
case the difference between direct and indirect or pseudodirect gaps 
is of the order of 0.1 eV only. 
In conclusion, the nature of the gaps is generally well described by our 
LDA calculations. 

Next, we consider the gap correction from the LDA. 
We note that both the Zn and Cd compounds have a gap correction 
$\Delta$ of about 1 eV ($\Delta$ varies from 0.83 eV for
ZnSiAs$_2$ to 1.13 eV for CdSiAs$_2$). This is
consistent with the gap corrections of GaP (1.2 eV),\cite{Zhu} GaAs 
\cite{Zhu,Godby} (0.9 eV),
and Zn-VI compounds \cite{Zakharov} (1.2--1.6 eV).

Other aspects of our band structures not discussed here are in
general  agreement with those of 
previous studies,\cite{Zung1,Zung2,Zung3,JaffeZunger}
for some of these compounds although there are some differences 
in the location of the Zn3d band.  
Thus, apart from the gap corrections, which for simplicity  we take 
from experiment and apply as a ``scissors'' shift to all conduction states, 
our band structures seem to describe the electronic structure of these systems 
rather well. 
In any rate, the optical response functions considered in this paper
are not very sensitive to the fine band structure details 
such as changes from direct to indirect at the level of 0.1 eV,
because they are derived from integrals over the whole Brillouin zone.
Also, the contributions of the Zn or Cd d-bands in the optical 
transitions determining $\chi^{(2)}$ are quite negligible
being far too deep although they do have an 
indirect effect on the lattice constants 
and the states near the valence band maximum via their hybridisation 
with the d-bands.

\subsection{Linear optical response and birefringence} \label{ss:lin}

For practical applications in SHG and OPO's an important quantity
is the birefringence because it enters in the phase-matching condition.
It can be calculated from the linear response functions from which
the anisotropy of the index of refraction is obtained.
Figure \ref{fig:eps_comp} shows the imaginary part of the dielectric
function (DF) $\varepsilon_2(\omega)$ for ZnGeP$_2$ and CdGeAs$_2$
in a wide energy region and for different light polarizations.
In this subsection we concentrate exclusively on these two materials
because of recent experimental  measurements of high accuracy
of frequency and temperature dependence of birefringence.
\cite{Zn-Ohm1,Zn-Ohm2,Cd-Ohm} In general, the $\varepsilon_2(\omega)$'s 
are rather similar in these materials, except for the difference in
the gap value. It consists mostly of one broad hump with a few additional
fine structures which depend on the light polarization. 

The birefringence is the difference between the
extraordinary and ordinary refraction indices, $\Delta n = n_e - n_o$,
where $n_e$ is the index of refraction for an electric field 
oriented along the c-axis and 
$n_o$ is the index of refraction for an electric field perpendicular to
the c-axis.
It is positive for both the materials ZnGeP$_2$ and CdGeAs$_2$.
In the low energy region ($\sim$ 0.1 eV) phonon absorption is starting to play
an important role, and this was clearly observed in the temperature
dependence of the birefringence in Refs.\onlinecite{Zn-Ohm1,Zn-Ohm2,Cd-Ohm}.
We cannot study the temperature dependence in the present publication.
However, when the temperature is changing between 14 K and 450 K,
the change of the birefringence at a given frequency is not larger than
10--15 \% of its value at room temperature, which is about the error
of our present estimation which is limited in its accuracy among others 
by the uncertainties in the gap corrections and
by the neglect of local field effects.

Figure \ref{fig:bir_comp}  shows $\Delta n$ in a wide energy region.
Of course, in practice the birefringence is important only
in the non-absorbing region, i.e., below the gap. The presense of
an absorption makes the use of the nonlinear crystal in OPO's or
frequency doubling device quite difficult or impossible. However,
such a curve is illustrative to show the general aspects of the difference
of the DF for different polarization. 
One may note that the general shape of the curves 
for ZnGeP$_2$ and CdGeAs$_2$ is rather similar, indicating the same frequency
regions where the $\epsilon_2(\omega)$ functions are enhanced or decreased 
in one polarization or the other. This, of course, is due to the 
similarities in their underlying band structures at the eV  scale.
Nevertheless, we can recognize a stronger negative peak 
around 2.5--3.0 eV in CdGeAs$_2$ than in ZnGeP$_2$ and a 
higher positive value in the low energy region.

The birefringence of ZnGeP$_2$ and CdGeAs$_2$ 
in the non-absorbing energy region 
together with experimental data is shown in Figures \ref{fig:bir_zgp}  
and \ref{fig:bir_cga}. As one could expect the calculated curve is
growing quadratically with energy (in the experiment $\Delta n$ starts
to grow again at low energies due to the phonon absorption).
The curve is continuing to grow with energy below the gap reaching
a non-monotonic behavior when the interband absorption processes
start.
For both ZnGeP$_2$ and CdGeAs$_2$ we predict fairly accurate values of the
birefringence to within about 10 \% in the middle of the gap region.
It correctly accounts for the fact that it is about 
twice larger in CdGeAs$_2$ than in ZnGeP$_2$.
For ZnGeP$_2$ the numerical values are in quite reasonable agreement
with experiment in all the non-absorbing region.
For CdGeAs$_2$ the agreement with measurements is good 
in a low energy region (0.1--0.3 eV) but 
at higher energies the experimental curve starts to grow
faster than the theoretical one. 
The reason for this disagreement can
be different, e.g.,
(i) the measurements could be prepared on a crystal which
is not perfect and contains impurities and cracks which increase
the scattering; (ii) the excitons (non included in our calculations)
could be important in this materials and give a significant 
contribution in refraction index near the band gap edge;
(iii) our calculations of the dielectric function for CdGeAs$_2$ 
might be not accurate enough.
Therefore, our calculations of birefringence should be considered as 
qualitative estimates only. Nevertheless, it is gratifying to see that even 
such an estimation exhibits the correct trends for the 
dependence of the birefringence on the 
chemical composition in these two chalcopyrites.
This shows that our calculations will be 
predictive for other chalcopyrites
and may be used for preliminary investigations 
of the phase-matching conditions
in these and related materials.

\subsection{Static limit of SHG} \label{ss:stat}

In this section we focus on the results for the  
SHG coefficients in the zero frequency limit.
For the material with the point group \={4}2$m$ there are only two
independent component of the SHG tensor, namely, the 123 and 312
components (1, 2, and 3 refer to the x, y, and z axes respectively,
which are chosen along the cubic axes).
In the static limit, these two components are becoming
equal according to Kleinman 
``permutation'' symmetry, \cite{Kleinman} which dictates additional
relations between tensorial components beyond the purely
crystallographic symmetry. 
For a non-ideal chalcopyrite structure (when $u$ $\neq$ 1/4) 
the point group symmetry is distorted and some additional components
of SHG (e.g., 311 or 131) appear. However, direct numerical calculations
show that 
these additional components are not larger than 3--5 \% of the main
123 component for the values of $u$ corresponding to real chalcopyrites. 
Therefore, we ignore them in the bulk of our results
and make all the calculations for the \={4}2$m$ point group.
For the considered ternary chalcopyrite semiconductors 
including the orbitals with angular moments $l_{max}$=3 in
the basis set changes the results by the factor of 5--10 \% only.
In most of our calculations we neglect them including 
orbitals with moments up to $l_{max}$=2.

In addition to the gaps discussed earlier, 
Table \ref{tab:shg} shows the values of the 
calculated and measured (where available)
SHG in the zero frequency limit in ABC$_2$ compounds
(A = Zn, Cd; B = Ge, Si; C = As, P).
We recall that all values were calculated using the  ``scissor'' 
approach with a rigid shift of the conduction bands chosen such that 
the minimum gap is adjusted to the experimental value. 

First, we discuss the trends with chemical element. 
One can see that substitution of P by As keeping the other elements 
fixed increases $\chi^{(2)}$ in all cases. This is also consistent 
with the values of GaAs (105 pm/V) and GaP (48 pm/V) 
calculated using the same approach (LDA+scissors) in Ref.\onlinecite{Our1}. 
Similarly, 
substitution of Si by Ge and Zn by Cd increases $\chi^{(2)}$.
Second, one may notice that this is correlated with the band gaps.
Substitution of As by P, or Ge by Si, or Cd by Zn, each 
increase the gap and decrease $\chi^{(2)}$. However, these 
qualitative trends cannot easily be turned into quantitative scaling 
laws. If one more carefully considers the ratios of the gaps 
and the ratios of $\chi^{(2)}$ for each of the above substitutions, 
no inverse power law between gaps and $\chi^{(2)}$ can be extracted
in the remaining four element combinations. 
For example, for substitution of As to P, the ratios of the gaps 
for ZnGe, ZnSi, CdGe and CdSi  are 1.8, 1.2, 3.0 and 1.4, respectively,
but the inverse ratio of the $\chi^{(2)}$ 's are 1.8, 1.7, 4.0 and 1.9.
On the basis of the theoretical expressions,\cite{Our1} one might
have guessed a scaling with the third power of the gap because 
three energy denominators occur. Clearly, such a strong scaling with 
the gap is not present. The reason for this is that there are 
several terms with positive and negative contributions which to 
a great extent cancel each other and secondly, the 
$\chi^{(2)}$ is not dominated by the terms involving the minimum gap
but arises from a {\bf k}- point summation over the whole Brillouin zone 
and a sum over several band-to-band terms.  Thus, no simple scaling with the 
gaps emerge.

The smallest value of the gap occurs
for CdGeAs$_2$ which also has the largest $\chi^{(2)}$.
From the above discussion of the trends, one
can certainly say that CdGeAs$_2$ has the mostly favorable
chemical combination of the elements
among the class of chalcopyrites considered here.
However, these arguments do not suffice to explain why the second harmonic
generation coefficient in CdGeAs$_2$ is so exceptionally large.

Before doing a more detailed analysis of the frequency dependent $\chi^{(2)}$
in the next section we first analyze the next two columns of the table
where we put separately interband versus intraband contributions
in SHG. The meaning of these terms needs perhaps some clarification.
In general, the formulas for second order response include many terms,
and its  division in  several groups called intra- or interband 
is rather conventional and depends somewhat on the formulation of the problem.
For instance, it appears at first to be different in the formulas using the 
length-gauge and the momentum gauge formulations. 
The most convincing decomposition based on a careful analysis of the 
underlying time dependent perturbation theory  
is the following \cite{Sipe,Aversa,Our1}:
the ``interband'' contribution corresponds to polarization loop where all
the three Green's function belong to different electronic bands,
while the ``intraband'' processes include
the modulation of the linear response by the intraband motion as well as
the modification of the intraband motion by the interband polarization
processes.

Analysis of the intraband and interband $\chi^{(2)}$ contributions
shows that, (i) for all the considered ABC$_2$ compounds 
the interband contribution is negative while intraband is positive,
therefore, these two terms work in opposite directions;
(ii) for all the materials except CdGeAs$_2$ and CdSiAs$_2$
both absolute values
of the intraband and interband terms are larger than
the resulting total SHG; (iii) the absolute value of the
intraband term is always larger than those of interband, i.e.,
the resulting total SHG is positive.
We also notice that CdGeAs$_2$ and CdSiAs$_2$ are exceptions which have 
an exceptionally small value of $\chi^{(2)}_{inter}$, 
i.e., the intraband term is dominating 
the SHG in the zero frequency limit. This term for CdGeAs$_2$ is nearly 
the same as in ZnGeAs$_2$. However, there is no interband 
contribution to compensate it. This explains why the second
response function is so large in this material.
The value of $\chi^{(2)}_{intra}$ in CdSiAs$_2$ is a few times smaller
than in CdGeAs$_2$ because of differences in the value of the gap.

Another point of interest is a comparison with the parent III-V compounds GaAs
and GaP. The values obtained at the same level of calculation\cite{Our1}
have already been given above.  Clearly, all phosphides considered here have a 
higher value of $\chi^{(2)}$ than GaP. They also have consistently 
lower gaps. For the arsenides, we also find the $\chi^{(2)}$ 
values to be consistently higher or equal in the chalcopyrites than 
in GaAs,  and to have lower gaps, except for ZnSiAs$_2$ which 
has a slightly higher gap and the same value of $\chi^{(2)}$ as GaAs.
The chalcopyrite type chemical substitution III-V to II-IV-V$_2$
thus clearly appears to be favorable for $\chi^{(2)}$. In addition, 
and even more important in practice, the anisotropy of the 
chalcopyrite crystal structure  allows for angular phase-matching while this 
is not possible in the cubic zincblende crystals.

The comparison of calculations with experiment is somewhat
complicated because the reported experimental results for these materials
differ considerably among each other. 
The last two columns of Table \ref{tab:shg} show experimental values of SHG
from the two different handbooks 
(Ref.\onlinecite{Landolt} and Ref.\onlinecite{Dmitriev}).
The values from Ref.\onlinecite{Dmitriev} are systematically higher than
those from Ref.\onlinecite{Landolt} by a factor of 1.3--1.4.
The reason of such a discrepancy is not completely clear. 
One possible explanation is that
most of the original values of SHG were measured relatively to quartz,
and the absolute value of second harmonic generation of pure quartz
has been remeasured a few times
(see Ref.\onlinecite{Roberts} for more recently recommended value
of SHG in quartz).
For ZnGeP$_2$, ZnSiAs$_2$, and CdGeP$_2$ our calculations are closer
to Ref.\onlinecite{Landolt} while for CdGeAs$_2$ they agree well
with Ref.\onlinecite{Dmitriev}. 
In general, we know that the ``scissors'' corrected values are overcorrected,
i.e., the actual value of SHG has to be higher \cite{Our1,Our2}.
Such an overcorrection is an intrinsic problem of the ``scissors'' approach.
With these cautions in mind, some of our values for the cases 
where no experimental values are available can be used as predictions.

\subsection{Frequency dependent SHG} \label{ss:freq}

In order to better understand the origin of the relative magnitudes 
of the intra- and interband contributions, we now consider the 
frequency dependent $\chi^{(2)}$ functions, or, more precisely,
their imaginary part from which the real part and 
in particular its static value
can be obtained by a Kramers-Kronig transformation.
Figures \ref{fig:shg_zgp} and \ref{fig:shg_cga} show the imaginary
part of the frequency
dependent $\chi^{(2)}(\omega)$ for the 123 and 312 components in
ZnGeP$_2$ and CdGeAs$_2$. First, it would be worth to mention that
the SHG curves for 123 and 312 components look very similar being 
different just in fine structure details.
This is not very surprising because in the initial zincblende material
which was used to derive the chalcopyrite material by chemical
substitutions these two components were equal at arbitrary frequency.
The shape of the curves is also qualitatively very similar in
the two materials. It consists of a broad hump with positive
SHG values between the half of the band gap and 1.5--2 eV, then
it changes sign and exhibits another broad dip between 2 and 3 eV.
At energies higher than 4 eV the imaginary part of SHG drops to
zero very fast. The amplitudes of both broad features are somewhat
higher in CdGeAs$_2$ than in ZnGeP$_2$. This is probably due to the smaller
value of the gap in the first material.
Im$\chi^{(2)}(\omega)$ for the other ternary pnictides ABC$_2$ considered
above are looking very similar, and we do not show them. 

As we see, the analysis of Im$\chi^{(2)}(\omega)$ does not
answer the question concerning the origin of the   
extremely large value of the zero frequency SHG
in CdGeAs$_2$.  To this end, it is more 
useful to analyze different contributions
to the total SHG curves.
Figures \ref{fig:ei_zgp} and \ref{fig:ei_cga} show the intraband and
interband contribution separately. 
For ZnGeP$_2$ both the interband and intraband contributions in the most
interesting energy region below 4 eV
(for the Kramers-Kronig integral which gives the zero frequency value)
do not change the sign
(the first is negative, the other is positive). This means that
both the inter- and intraband contributions to the static SHG are large
and have opposite signs.
In CdGeAs$_2$ the situation is different.
While the intraband term behaves in the similar way to those in
ZnGeP$_2$ (does not change the sign below 4 eV), the interband term
is positive between 0.3 and 1.1 eV and negative between 1.1 and 3 eV.
Thus, the low energy region (below 4 eV) of the 
interband term by itself already exhibits a good 
deal of compensation between positive and negative values when the 
integral appearing in the Kramers-Kronig transformation is calculated.
There is no simple qualitative explanation of the fact why
the resulting Kramers-Kronig integral which gives the zero frequency
value of the interband contribution to SHG is as small 
as the numerical calculations show: less than
0.5 \% of the intraband term in our calculations, see the previous
section). This appears to be a rather coincidental 
interplay between different interband
contributions. The important fact is that the contributions
from different groups of interband transitions in this material
have different sign, and they compensate each other in the Kramers-Kronig
integral. 

This example shows that a small value of the gap may be favorable for
but is clearly not sufficient or the dominant contributing factor
to obtain a  large value of SHG.
The value of the SHG results from a delicate balance between different 
contributions too complicated to capture in a simple scaling law with the gap.
Unfortunately, we cannot presently further reduce this question 
to a simple explanation in terms of particular band-to-band transitions.
Nevertheless, it shows that the entire band structure
matters, not simply the gaps.  Therefore, there appears, at least in principle,
to be an ample room for optimization 
of $\chi^{(2)}$ by band structure engineering.
Of course, we cannot control the signs of different interband
contributions in a definite bulk chemical compound at will even with 
quite elaborate chemical substitutions.  
However, further opportunities to modify the band structures 
could exist for 
semiconducting heterostructures, such as 
multiple quantum wells, superlattices, layered materials, quantum dots, etc.
In these structures the energies of the transitions between
different levels and their matrix elements
can to some extent be controlled by geometrical parameters 
of the structure.
For example, it has already been shown that
the asymmetric quantum wells can display
very large values of second order optical response (see, e.g.,
Refs.\onlinecite{Capasso1,Capasso2}).
However, the situation in these systems is complicated, and
theoretical studies show that excitonic effects play an important
role,\cite{Agran,Tsang,Voon} in addition to the single electron part
considered in our work.
Nevertheless, the idea of parametrically controlled 
heterostructures with large 
$\chi^{(2)}$ 
is a challenging task which we hope to pursue in future work
because of the practical importance of obtaining materials 
with higher $\chi^{(2)}$ values.

\subsection{Structure dependence} \label{ss:str}

As we already mentioned above, the distortions from the ideal chalcopyrite
structure can be neglected in calculations of optical spectra and SHG
only if they are small enough as in the case of ZnGeP$_2$.
Figure \ref{fig:bnd1_zgp} compares the electronic bands for
ZnGeP$_2$ along the $Z-\Gamma$ and $\Gamma-X$ lines calculated in
two different crystal structures, namely, the
real experimental structure 
(with the parameters from Table \ref{tab:str}), and the ideal chalcopyrite
structure with the same lattice constant $a$.
The band structures look very similar being different only in some
fine details
(e.g., the splitting of the upper valence band at the $\Gamma$- point 
is different in the two
cases due to different crystal fields).
Both crystals have indirect ($\Gamma-X$) band gaps of nearly the same value.
The zero-frequency SHG values calculated in LDA 
are also very close (102 pm/V for the
experimental structure and 106 pm/V for ideal chalcopyrite,
i.e., the difference between them is only about 4 \%).

In materials with larger distortions the situation is different.
One has to study how the deviations of both the parameters $\eta$ and $u$ 
from those of the ideal chalcopyrite structure ($\eta$ = 2, $u$ = 1/4)
affect both the band structure and optical constants.
Figure \ref{fig:bnd1_csp} shows how the value of the parameter $u$ affects
the electronic bands in CdSiP$_2$ (this compound is the most illustrative
because the deviations of both $\eta$ and $u$ are the largest ones when
compared with other materials of the considered group). 
Except for some fine details, the effect of $u$ can be considered as a rigid
shift upwards (by about 0.4 eV) of the low conduction bands.
The character of the gap (pseudodirect) does not change with $u$.
The behavior of the band structure is completely different when we change the
parameter $\eta$ (Fig. \ref{fig:bnd2_csp}). Namely, when the value
of $\eta$ is approaching 2, the two lowest conduction bands are changing
their positions, and the band gap is becoming direct as opposed to indirect.
This illustrates the differences of the behavior of the electronic
spectra with changes in $\eta$ and $u$.

For the other materials the situation is quite similar. The change of
$u$ produces nearly rigid shift of the few lowest conduction bands
while the changes of the parameter $\eta$ can mix these bands and
change the character of the band gap. Moreover, the value of the rigid
shift is nearly linear in $(u-1/4)$ and depends very weakly
on the material. Thie means that the deformation
potential associated with this structural change is almost constant 
in this class of compounds. Why this is so 
is presently not clear although it is not entirely unexpected 
since, in general, these materials are of course
all closely related to GaAs or GaP. 

In order to study how these changes affect the optical constants we calculated
the zero frequency SHG coefficients at different values of $\eta$ and $u$.
For the band structures of CdSiP$_2$ shown in Figures \ref{fig:bnd1_csp} and
\ref{fig:bnd2_csp} this calculation gives $\chi^{(2)}$ = 73 pm/V for
the real experimental structure, $\chi^{(2)}$ = 90 pm/V for
$\eta$ = 1.836 and $u$ = 1/4, and $\chi^{(2)}$ = 113 pm/V for
the ideal chalcopyrite structure. The value of the lattice constant
$a$ and the value of the gap correction used 
was the same in all the three cases.
These calculations illustrate qualitatively
how SHG depends on distortions of the ideal
chalcopyrite structure. In the considered case both the deviations
of $\eta$ and $u$ tend to decrease the value of SHG.

Another interesting question
is to study whether or not the 
chalcopyrite type of ordering of 
the cations plays a 
special role in high $\chi^{(2)}$ values in these materials. 
For this purpose we considered an alternative crystal structure,
a (001) oriented $1+1$ superlattice AC+BC, which still
maintains the local tetrahedral environment
of the chalcopyrite where each anion is
surrounded by two cations of each type but which nevertheless 
exhibits a different ordering vector.
This superlattice structure has the same overall
chemical formula and stoichiometry as the chalcopyrite structure.

Figure \ref{fig:bnd2_zgp} shows the LDA band structures of the $1+1$
ZnGeP$_2$ 
superlattice along the same directions of the Brillouin zone as in
Fig. \ref{fig:bnd1_zgp}.
The real cell for the superlattice is a
tetrahedral one which is twice smaller than the cell for the
chalcopyrite structure. 
We also calculated the electronic structure of $1+1$ superlattices 
for other ABC$_2$ materials considered before. However, in some materials
which exhibit a small LDA gap in the chalcopyrite structure,
the gap for the $1+1$ superlattice becomes negative. 
Here we show the results for ZnGeP$_2$ only in order to illustrate
how the band structure changes with modifications in the crystal
structure.
The gap is much smaller in the $1+1$ structure ($E_g^{LDA}$ = 0.3 eV) 
than in the chalcopyrite ($E_g^{LDA}$ = 1.16 eV).
As a result, the zero frequency SHG is about 190 pm/V for the
superlattice, i.e., about twice higher than 
SHG in the chalcopyrite ZnGeP$_2$
(we used the same value of the gap correction ($\Delta$ = 0.89 eV)
for the superlattice structure which has been used before for the
chalcopyrite ZnGeP$_2$).

While this is at first sight promising for obtaining higher
$\chi^{(2)}$ we caution that this structure is unlikely to be stable.
The atomic sizes of cations are different,
and one can expect an occurance of alternatingly compressive and
tensile biaxial strain in the cation layers.
The calculations show that the difference of
the total energy per chemical unit between the superlattice
and the chalcopyrite structure is $\Delta E_{tot}$ = 0.204 eV, 
i.e., is rather large.
This confirms 
our hypothesis that the alternative structure is not
stable with respect to different structural rearrangements 
minimizing the total energy.

\section{Conclusions} \label{sec:con}

The linear and nonlinear optical properties for important group of
chalcopyrites ABC$_2$ (A= Zn, Cd; B= Ge, Si; C= As, P) have been calculated
over a wide energy range. 
We studied any possible combination of A, B, and C.
This allowed us to study
the trends in the second order optical response with chemical composition.
The results for the zero-frequency limit of SHG
are in good agreement with available experimental results.
The calculated birefringence for ZnGeP$_2$ and CdGeAs$_2$ also shows
a fair agreement with recent experimental data in the energy
region corresponding to the middle of the gap.
For all the considered compounds the second harmonic generation
coefficient $\chi^{(2)}$ is of the order of 100--200 pm/V, i.e., of 
the order of and in fact larger than the SHG 
in the initial zincblende material (GaAs or GaP)
from which these compounds were created by chemical substitutions.
The only exception is CdGeAs$_2$, which has much higher nonlinear
response ($\sim$ 500 pm/V) than all the other materials from this group.
This value of the SHG cannot be explained only by a small value of
the band gap. It appears as a result of a very delicate balance between
different terms which contribute to second order response.

The results of the calculations are rather stable with respect to
small structural modifications. They do not change much with 
distortions of the ideal chalcopyrite structure in materials
in which $\eta$ and $u$ are close to those in ideal chalcopyrite
structure. However, when these deviations are large (like
in CdSiP$_2$) the gaps as well as SHG's can be strongly affected by the values
of $\eta$ and $u$ chosen in calculations, and the ideal
structure does not describe a real experimental
situation correctly.
The LDA calculations
for the (001)- oriented $1+1$ superlattice which has the same overall
chemical formula and stoichiometry as the chalcopyrite structure 
exhibits a much smaller LDA band gap than the chalcopyrite structure
and larger values of the SHG
coefficients. However, this structure has a very high total energy
which is probably related to large strains in the cation layers. 
These strains should make the structure unstable with respect to
different structural rearrangements.

In conclusion, we can certainly say that a small value of the 
gap is favorable for larger SHG. However, this 
is by no means sufficient for quantitative predictions of the 
$\chi^{(2)}$ values from the known values of the gaps. 
Within this family of compounds all calculated in essentially 
the same approximation, no clear scaling of calculated $\chi^{(2)}$ values 
with a power law of the gap was obtained.  
Because of the experimental uncertainties in determining $\chi^{(2)}$
values for compounds with different degree of crystalline ideality 
and different optimization of phase-matching conditions, we consider
this test as a more reliable test of such scaling laws than those based 
on experimental values.  
The origin of $\chi^{(2)}$ in terms of the underlying 
band structure is  clearly too complex for such a simple minded
extrapolation to be valid. 
One has to take into account all the contributing terms
and analyze them carefully. Even then, gaining understanding 
is far from easy and we were only partially successful at 
unraveling the origin of the high value of $\chi^{(2)}$ for CdGeAs$_2$.
Nevertheless, we hope that our present analysis in terms
of the intra- and interband terms provides at least a first
step towards such understanding. 
Both the intra- and interband terms are generally found to be 
decreasing with increases 
of the gap because both term contain some 
power of the interband transition energy in the denominator. 
Even, so, they do not simply scale with the gap and their sum 
which is the total value of $\chi^{(2)}$ can either increase
or decrease depending on the degree of compensation between  the
two contributions which was found to depend sensitively 
on details of the calculations, for example, the ``non-ideality''
of the parameter $u$. 
Thus, only a very complicated interplay
between these different terms is forming the total value of
the nonlinear optical response function.
A detailed analysis of these contributions appears to be necessary 
to understand and predict confidently 
the expected NLO response for new materials.

\acknowledgements
We wish to thank Prof. B. Segall for useful discussions.
This work was supported by NSF (DMR95-29376).

%
%

\newpage

\begin{table}
\caption{
Experimental values of lattice parameters $a, \eta$, and $u$ used in
the present calculations. 
}
\label{tab:str}
\begin{tabular}{lcccl}
 & & & & \\
Compound      &  $a$ (a.u.)  &  $\eta$  &  $u$  &  Reference   \\
 & & & & \\
\tableline
 & & & & \\
ZnGeP$_2$     & 10.317 & 1.970  &  -      
&  Ref.\protect\onlinecite{Wyckoff} \\
	      & 10.324 & 1.965  &  0.2582 
&  Ref.\protect\onlinecite{Landolt2}\\ 
 & & & & \\
ZnGeAs$_2$    & 10.716 & 1.966  &  0.2585    
&  Ref.\protect\onlinecite{Landolt2} \\
 & & & & \\
ZnSiP$_2$     & 10.200 & 1.934  &  -      
&  Ref.\protect\onlinecite{Wyckoff} \\
	      & 10.204 & 1.933  &  0.2691 
&  Ref.\protect\onlinecite{Landolt2}\\
 & & & & \\
ZnSiAs$_2$    & 10.598 & 1.941  &  -      
&  Ref.\protect\onlinecite{Wyckoff} \\
	      & 10.593 & 1.940  &  0.2658 
&  Ref.\protect\onlinecite{Landolt2}\\
 & & & & \\
CdGeP$_2$     & 10.843 & 1.878  &  -      
&  Ref.\protect\onlinecite{Wyckoff} \\
              & 10.847 & 1.878  &  0.2819       
&  Ref.\protect\onlinecite{Landolt2} \\
 & & & & \\
CdGeAs$_2$    & 11.229 & 1.889  &  0.285  
&  Ref.\protect\onlinecite{Wyckoff} \\
              & 11.230 & 1.888  &  0.279  
&  Ref.\protect\onlinecite{Landolt2} \\
 & & & & \\
CdSiP$_2$     & 10.731 & 1.836  &  0.2967
&  Ref.\protect\onlinecite{Landolt2} \\
 & & & & \\
CdSiAs$_2$    & 11.121 & 1.849  &  0.2893
&  Ref.\protect\onlinecite{Landolt2} \\ 
 & & & & \\
\end{tabular}
\end{table}

\begin{table}
\caption{
Experimental (Ref.\protect\onlinecite{Landolt2})
and LDA values of the energy gap, ($E_g^{expt}$ and $E_g^{LDA}$), 
their difference ($\Delta$),
calculated and measured values of the static $\chi^{(2)}$ 
and its decomposition in inter- and intraband contributions 
for different ternary semiconductors ABC$_2$ (in pm/V).
The values of $\chi^{(2)}$ are compared with experimental
data from Refs. \protect\onlinecite{Dmitriev} and 
\protect\onlinecite{Landolt}.
We use the symbols (d), (pd) and (i) for direct, pseudodirect 
and indirect gaps
respectively. 
}
\label{tab:shg}
\begin{tabular}{l|ccc|ccc|c|c}
 \\
Compound  & $E_g^{expt}$ (eV) & $E_g^{LDA}$ (eV) & $\Delta$ (eV)   
&  $\chi^{(2)}_{total}$ & $\chi^{(2)}_{inter}$ & $\chi^{(2)}_{intra}$
&  $\chi^{(2)}$ (Ref.\protect\onlinecite{Landolt}) 
&  $\chi^{(2)}$ (Ref.\protect\onlinecite{Dmitriev}) \\
& & & & & & & &   \\
\tableline
 & & & & & & & & \\
ZnGeP$_2$  & 2.05 (pd) & 1.16 (i) & 0.89 & 102 & -231 & 333 & 111 & 150 \\
  & & & & & & & & \\
ZnGeAs$_2$ & 1.15 (d) & 0.13 (d)  & 1.02 & 185 & -328 & 513 & -   & -   \\
 & & & & & & & & \\
ZnSiP$_2$  & 2.07 (pd) & 1.22 (pd) & 0.85 & 61 & -220 & 281 & -   & -   \\
 & & & & & & & & \\
ZnSiAs$_2$ & 1.74 (pd) & 0.91 (d)  & 0.83 & 105 & -210 & 315 & 109 & 146 \\
 & & & & & & & &\\
CdGeP$_2$  & 1.72 (d) & 0.75 (d)   & 0.97 & 127 & -151 & 278 & 162 & 218 \\
 & & & & & & & &  \\
CdGeAs$_2$ & 0.57 (d)  & -0.44 (d) & 1.01 & 506 & -2   & 508 & 351 & 472 \\
 & & & & & & & &  \\
CdSiP$_2$  & 2.2--2.45 (pd)       &  1.19 (pd) 
& 1.01 & 73 & -139 & 212 & -   & -   \\
 & & & & & & & & \\
CdSiAs$_2$ & 1.55 (d)    & 0.42 (d)  & 1.13 & 139 & -21 & 159 & -   & -   \\ 
 & & & & & & & &  \\
\end{tabular}
\end{table}

\begin{figure}
\caption{
Calculated $\varepsilon_2(\omega)$ for ZnGeP$_2$ and CdGeAs$_2$ 
(shifted down for convenience).
The solid line corresponds to the polarization {\bf E} $\perp$ {\bf c},
the dotted line corresponds to {\bf E} $\parallel$ {\bf c}. 
}
\label{fig:eps_comp}
\end{figure}

\begin{figure}
\caption{
The calculated $\Delta n$ for ZnGeP$_2$ (solid line) and
CdGeAs$_2$ (dotted line) in a wide energy region. 
}
\label{fig:bir_comp}
\end{figure}

\begin{figure}
\caption{Calculated birefringence for ZnGeP$_2$
(solid line) together with experimental
data of Ref.\protect\onlinecite{Zn-Ohm2} at room temperature.
The open circles and filled
circles correspond to measurements on two different ZnGeP$_2$ samples.
}
\label{fig:bir_zgp}
\end{figure}

\begin{figure}
\caption{Calculated birefringence for CdGeAs$_2$ (solid line) together
with room temperature measurements of Ref.\protect\onlinecite{Cd-Ohm} 
(filled circles).
}
\label{fig:bir_cga}
\end{figure}

\begin{figure}
\caption{
Calculated 123 (solid line) and 312 (dotted line) components
of the imaginary part of SHG for ZnGeP$_2$.
}
\label{fig:shg_zgp}
\end{figure}

\begin{figure}
\caption{
Calculated 123 (solid line) and 312 (dotted line) components
of the imaginary part of SHG for CdGeAs$_2$.
}
\label{fig:shg_cga}
\end{figure}

\begin{figure}
\caption{
Calculated interband (solid line) and intraband (dotted line) contributions
to Im$\chi^{(2)}(\omega)$ for ZnGeP$_2$ (123 component).
}
\label{fig:ei_zgp}
\end{figure}

\begin{figure}
\caption{
Calculated interband (solid line) and intraband (dotted line) contributions
to Im$\chi^{(2)}(\omega)$ for CdGeAs$_2$ (123 component).
}
\label{fig:ei_cga}
\end{figure}

\begin{figure}
\caption{
Electronic bands of ZnGeP$_2$ along the $Z-\Gamma$ and $\Gamma-X$ lines in
the Brillouin zone ($Z= 2\pi/\eta a \cdot(0,0,1)$; $\Gamma= (0,0,0)$;
$X= 2\pi/a\cdot(1/2,1/2,0)$ ) for the experimental crystal structure
(solid lines) and for the ideal chalcopyrite structure ($u$=1/4,
$\eta$=2) with the same lattice constant $a$ (dotted lines).
Both the structures are shown in LDA (without gap corrections).
The energy in both cases is counted from the top of the valence band.
}
\label{fig:bnd1_zgp}
\end{figure}

\begin{figure}
\caption{
The LDA electronic bands of CdSiP$_2$
for the experimental crystal structure ($\eta$ = 1.836, $u$ = 0.2967,
solid lines) compared with bands for the structure with the same
$a$ and $\eta$ but with the value of $u$
for the ideal chalcopyrite structure ($u$=1/4, dashed lines).
}
\label{fig:bnd1_csp}
\end{figure}

\begin{figure}
\caption{
The LDA electronic bands of CdSiP$_2$
for the crystal structure with ($\eta$ = 2, $u$ = 1/4,
solid lines) compared with bands for the ideal chalcopyrite
structure ($\eta$ = 2, $u$=1/4, dashed lines) with the same lattice constant
$a$.
}
\label{fig:bnd2_csp}
\end{figure}

\begin{figure}
\caption{
The LDA electronic bands
for the $1+1$ alternative ZnGeP$_2$ structure.
}
\label{fig:bnd2_zgp}
\end{figure}

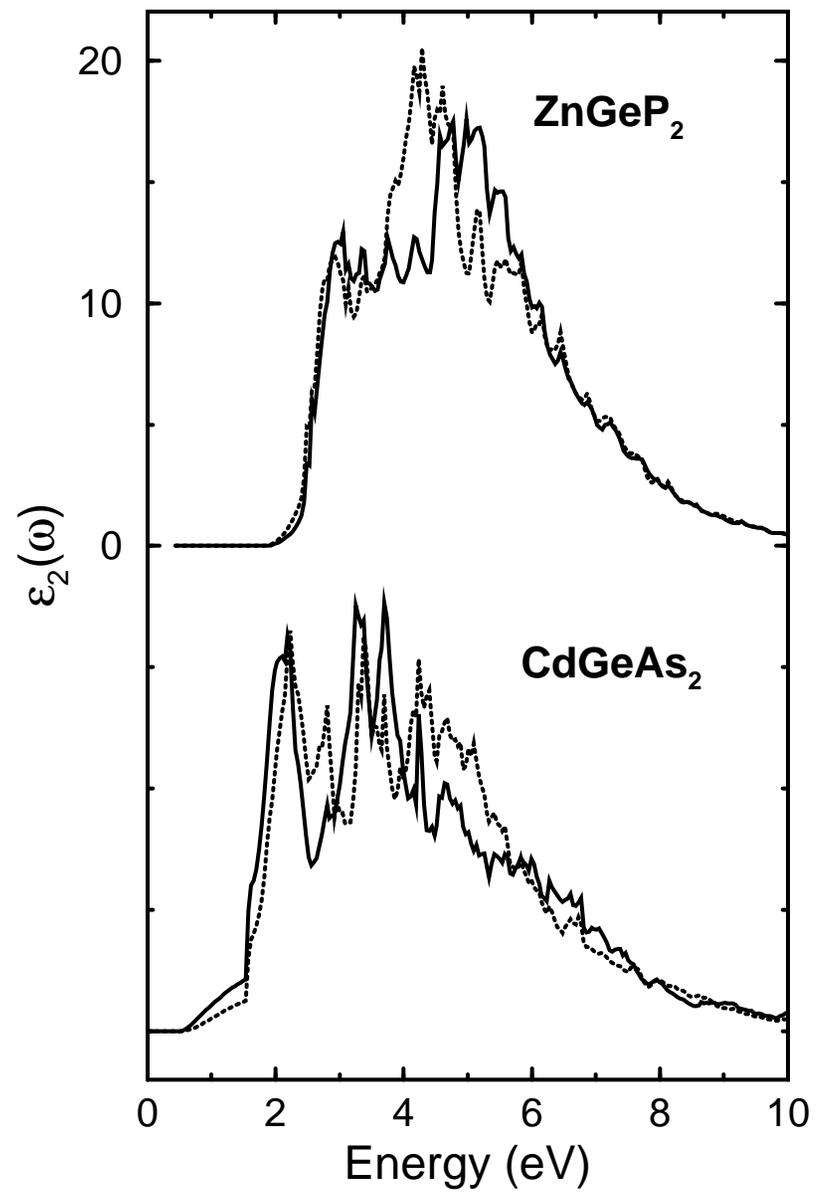
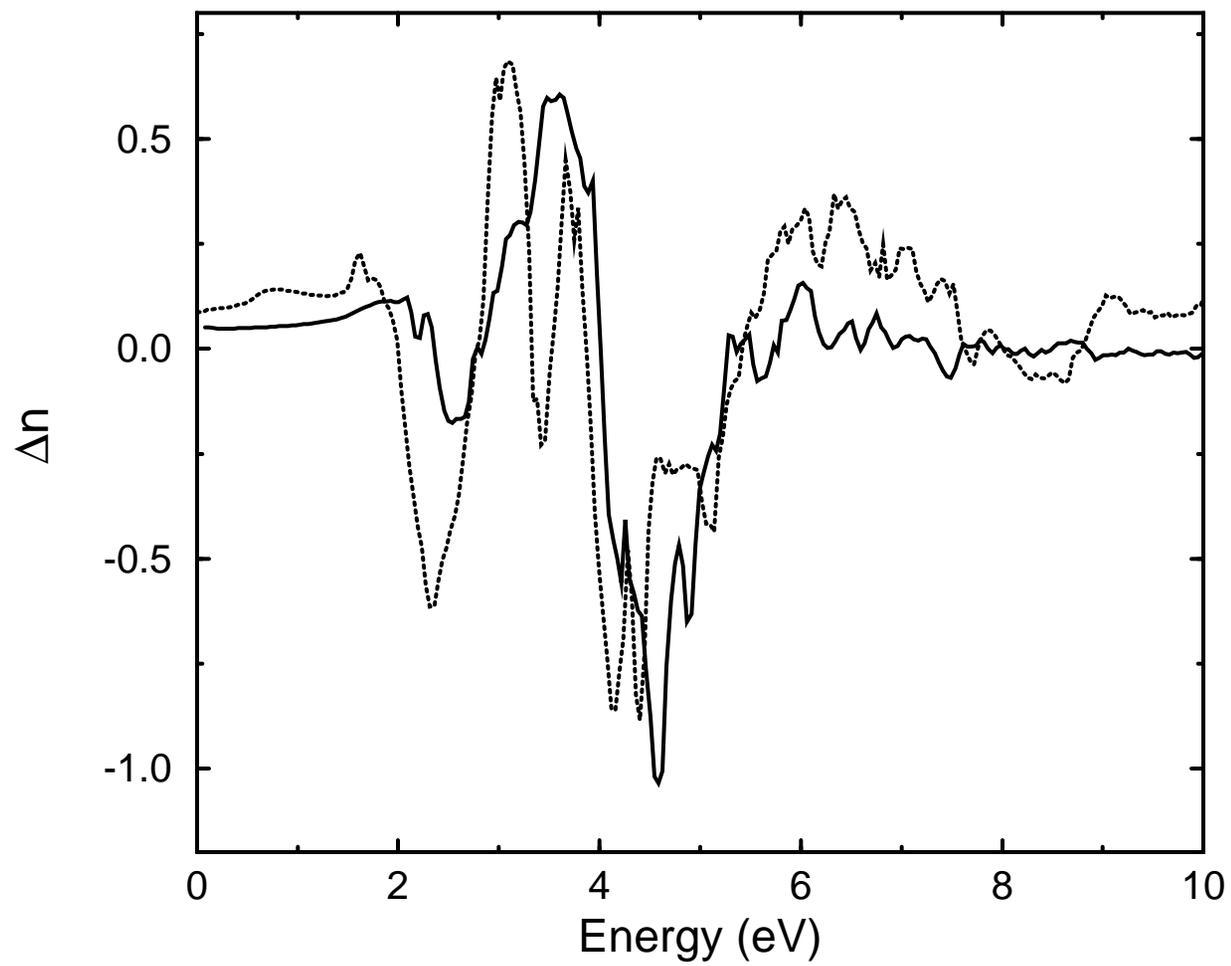
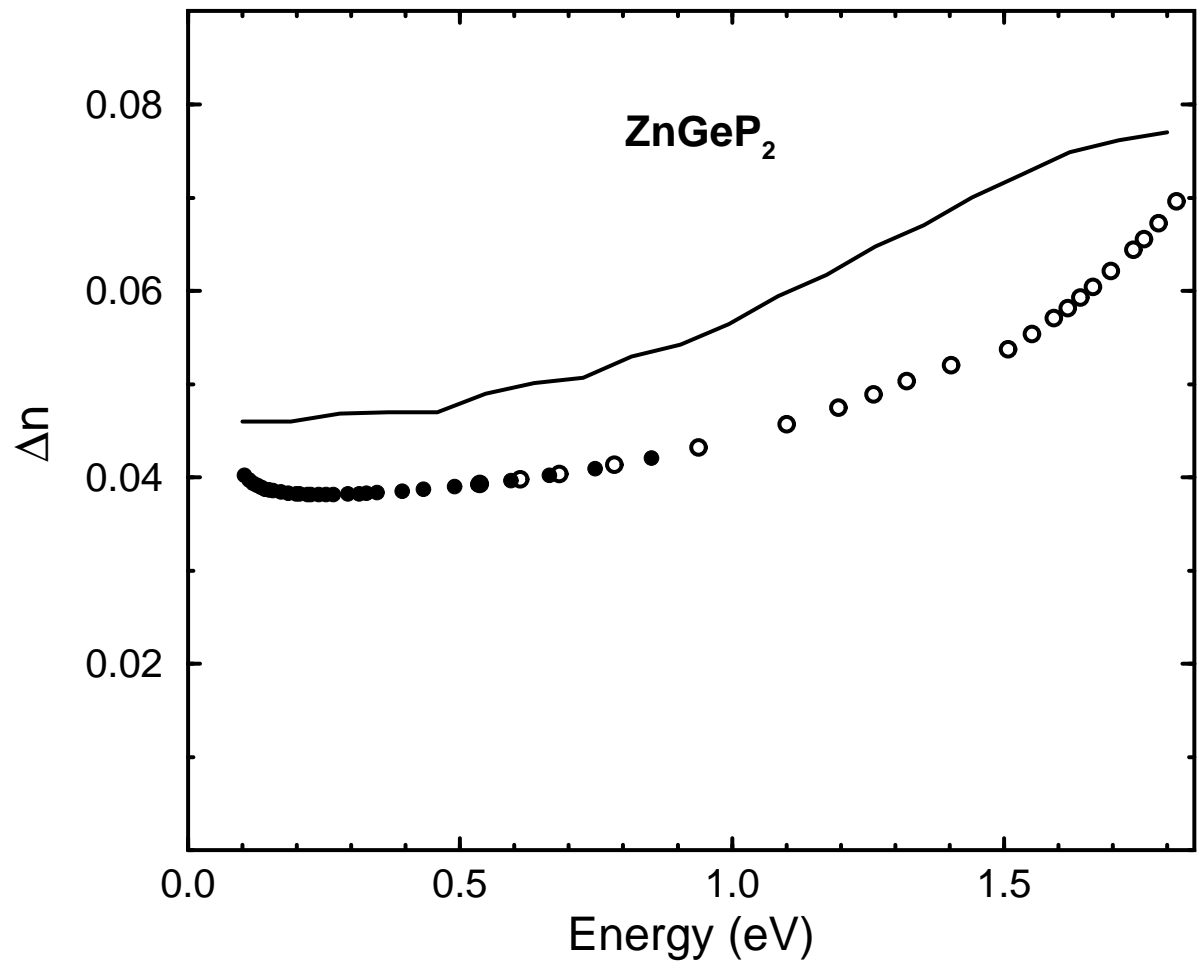
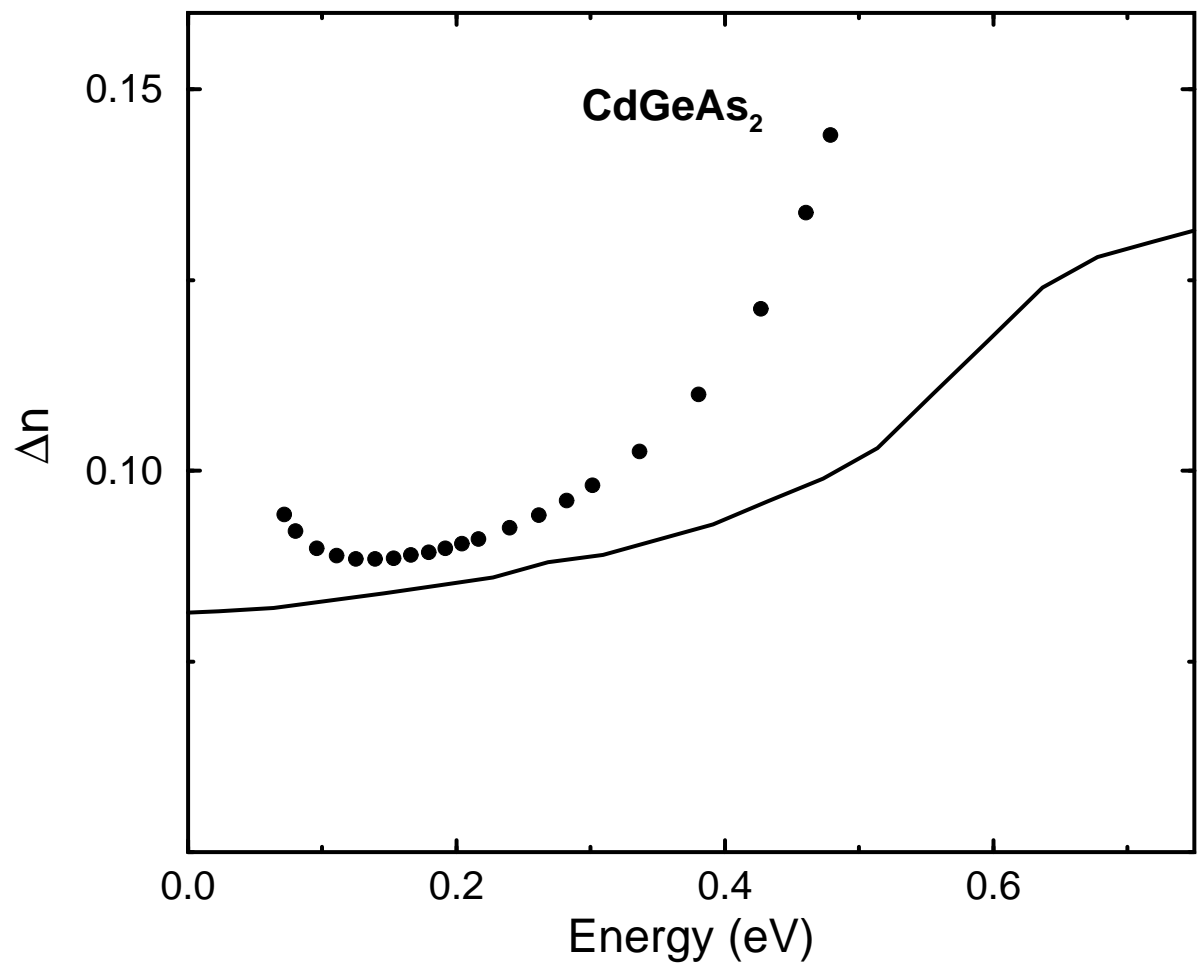
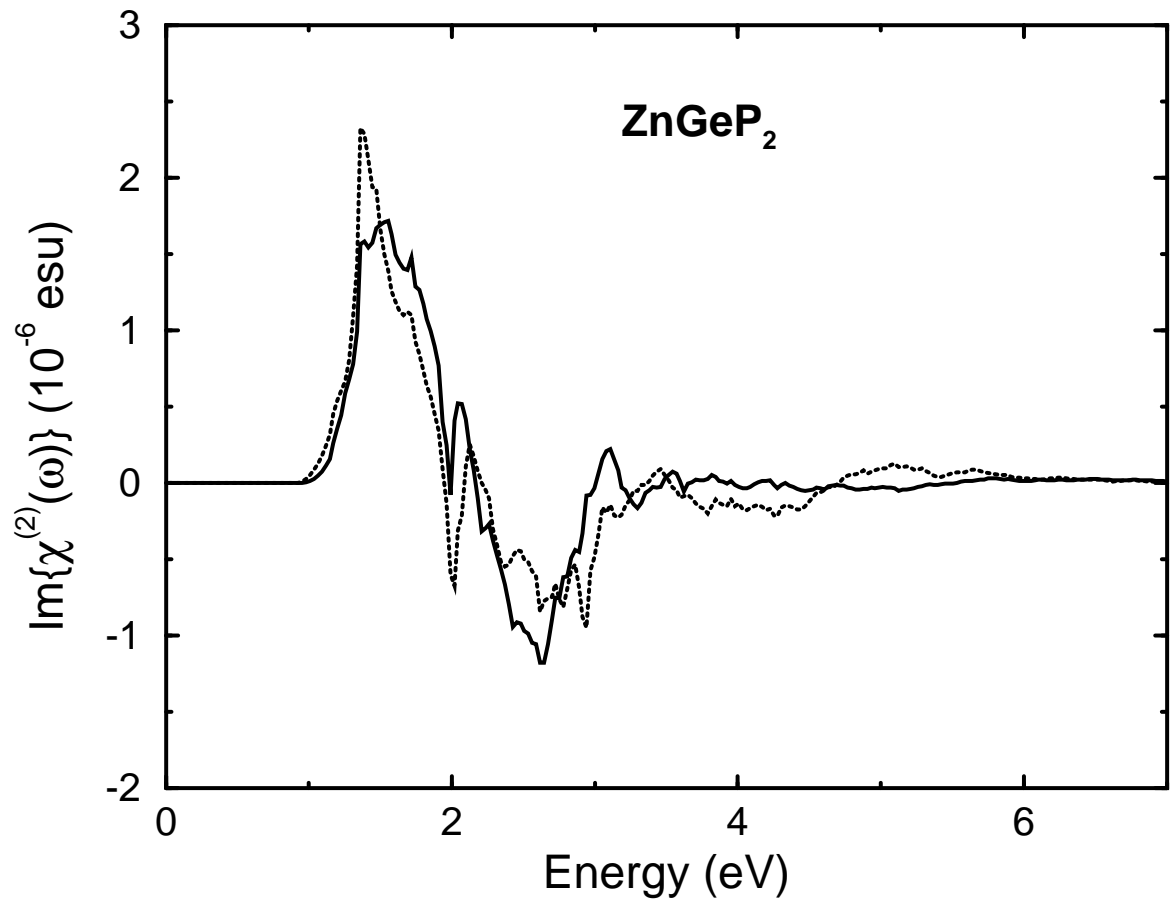
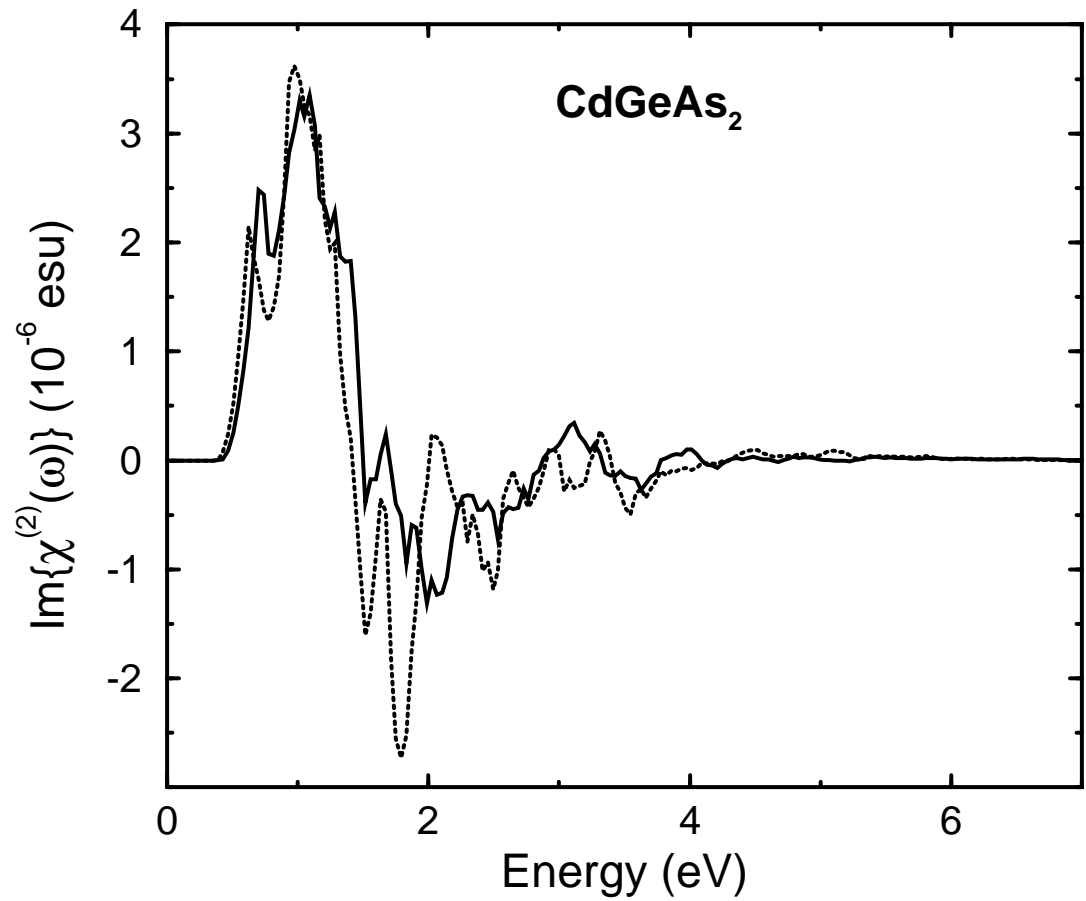
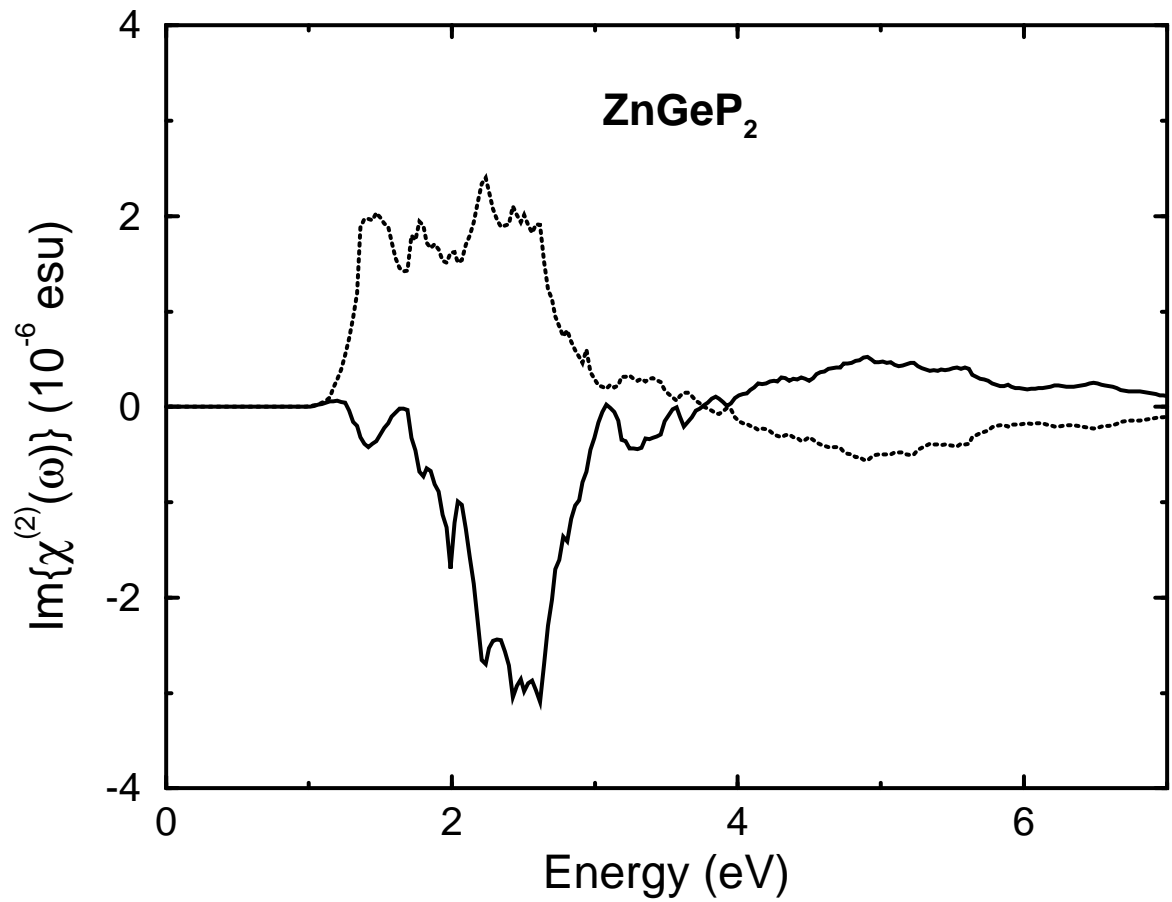
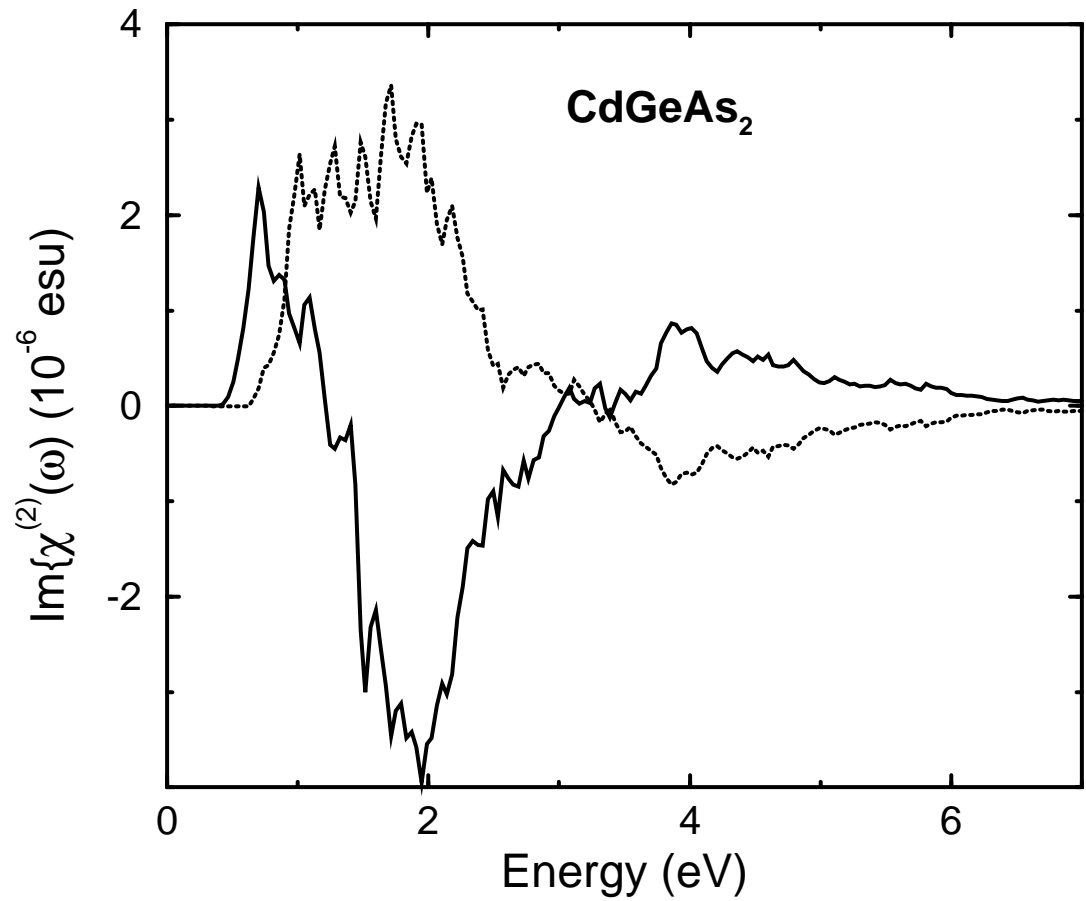
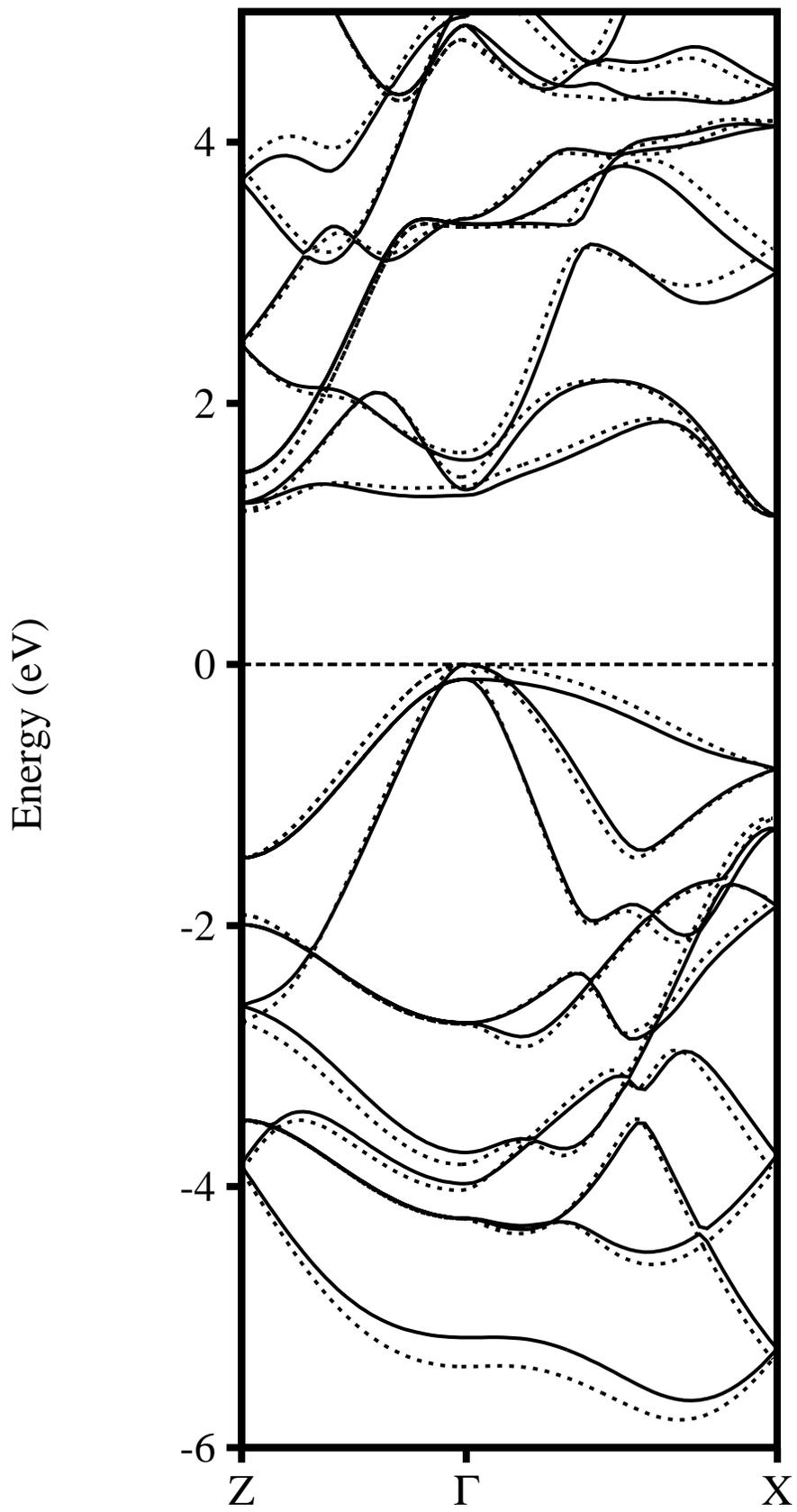
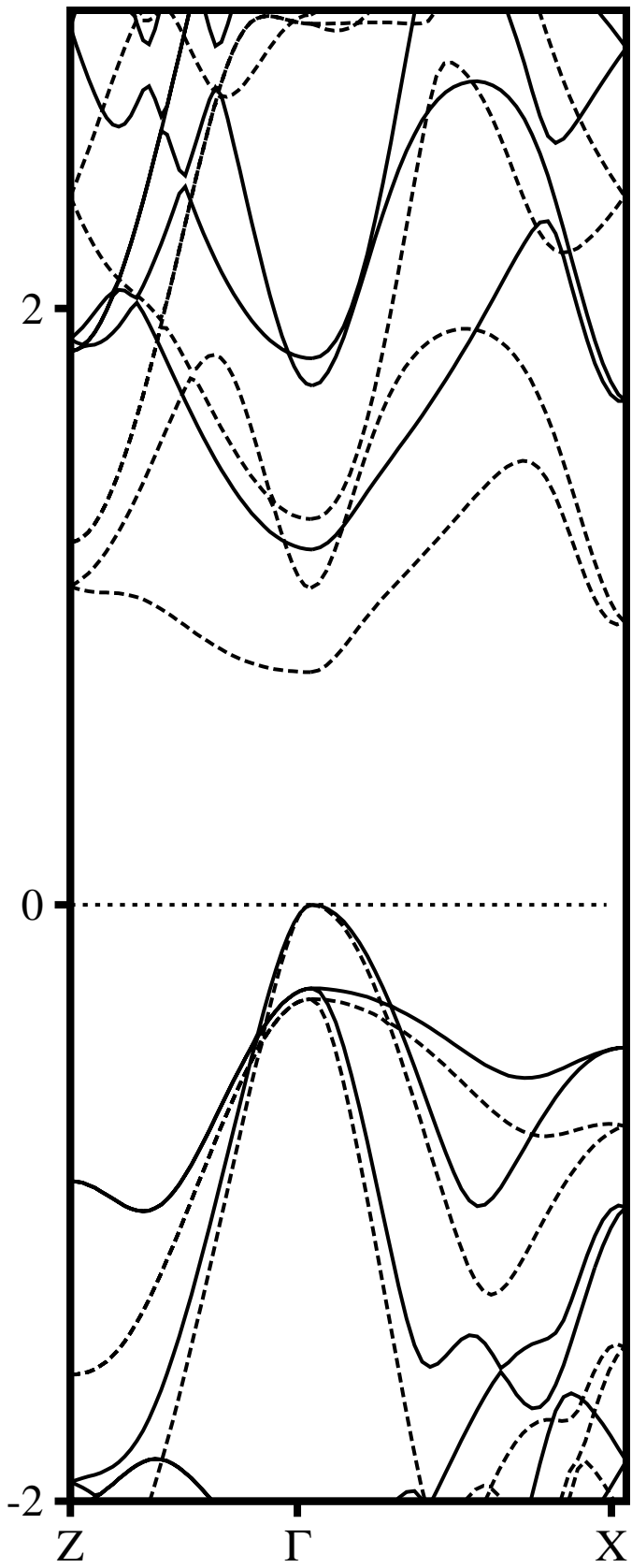
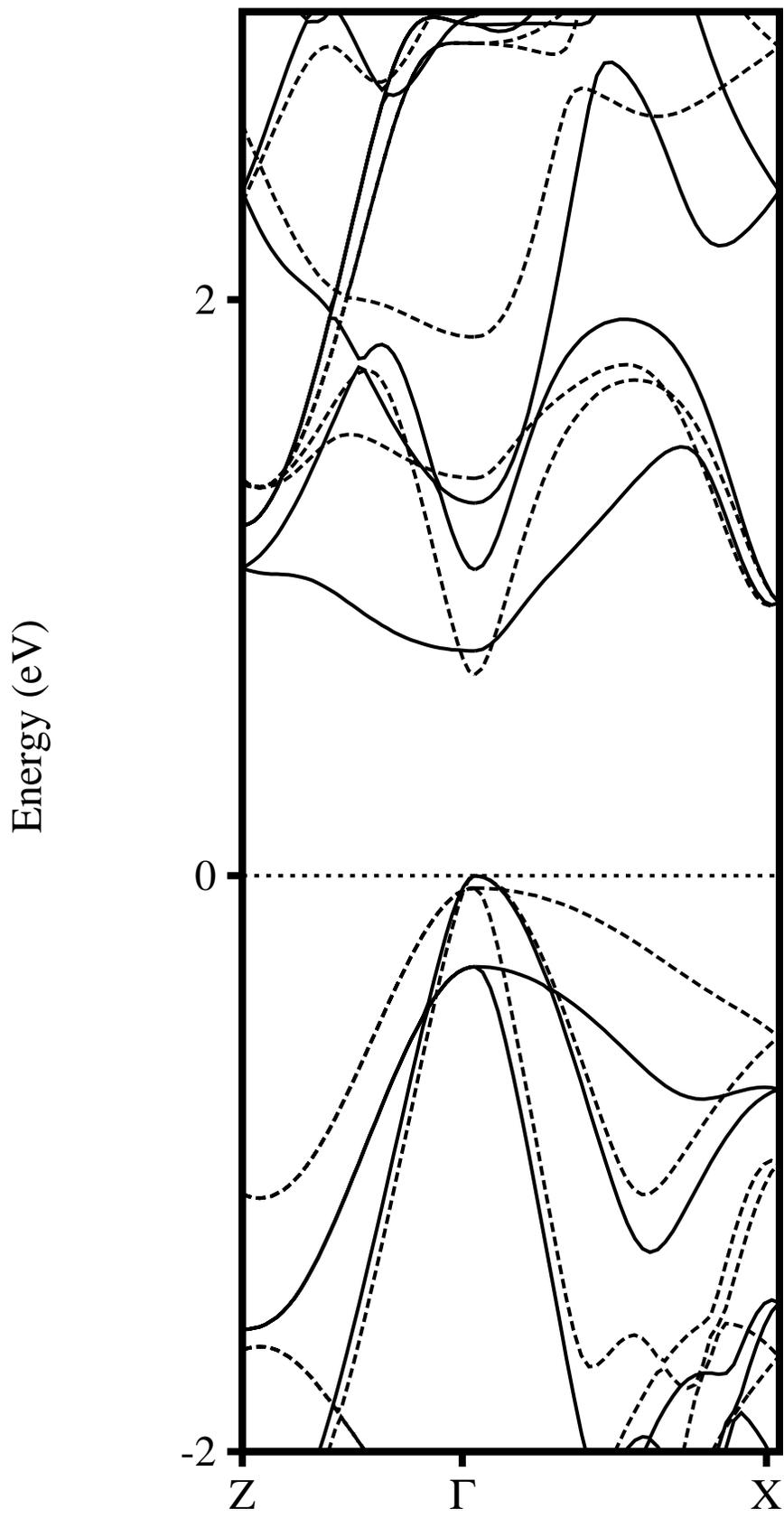
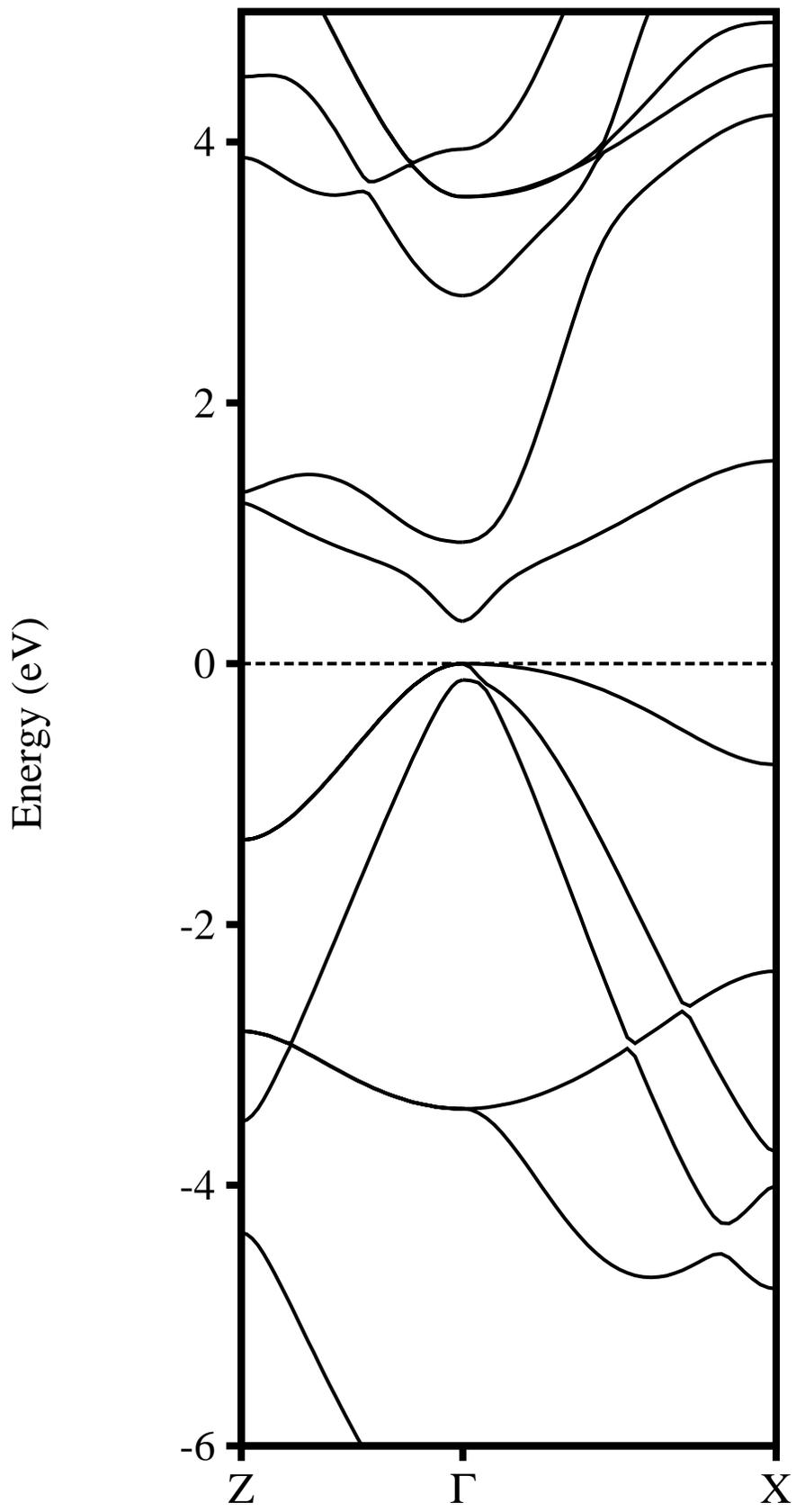
\end{document}